\documentclass[12pt,onecolumn]{article}

\usepackage[utf8]{inputenc} % Allow utf-8 input
\usepackage[T1]{fontenc}    % Use 8-bit T1 fonts
\usepackage{hyperref}       % Hyperlinks
\usepackage{url}            % Simple URL typesetting
\usepackage{booktabs}       % Professional-quality tables
\usepackage{amsfonts}       % Blackboard math symbols
\usepackage{nicefrac}       % Compact symbols for 1/2, etc.
\usepackage{microtype}      % Microtypography
\usepackage{lipsum}         % For generating placeholder text
\usepackage{graphicx}       % For including figures
\usepackage{amsmath}        % Advanced math typesetting
\usepackage{setspace}       % Custom spacing control
\usepackage{natbib}

\usepackage{geometry}
\newcommand{\kruskal}[1]{\left[\!\left[#1\right]\!\right]}
\newcommand{\bigsum}{\mathop{\vcenter{\hbox{\scalebox{2.0}{$\displaystyle\sum$}}}}}

\title{Holistic Multivariance Decomposition: Adapting Mode Interrelations in Low-Rank Tensor Approximations}

\author{
 \begin{tabular}{c}
    Süha Tuna\thanks{Corresponding author.} \\
    \small Informatics Institute, Department of Computational Science and Engineering,  \\
    \small \.Istanbul Technical University, \.Istanbul 34469, T\"urkiye \\
    \small \texttt{suhatuna@itu.edu.tr}
  \end{tabular}
}

\date{}

\begin{document}

\maketitle

\begin{abstract}
Low-rank tensor approximation is a foundational tool for multidimensional data analysis in scientific computing, classically dominated by Tucker and Canonical Polyadic (CP) decompositions. While widely adopted, these standard approximation schemes represent data as sums of rank-$1$ tensors formed via mode-wise outer products. This inherent mathematical structure captures the independent variations of individual modes but systematically neglects the mutual interactions and coupled dimensional interdependencies natively embedded within the tensor. To overcome this structural limitation, we introduce the Holistic Multivariance Decomposition (HMD) framework. HMD provides a novel tensor decomposition algorithm that explicitly models both isolated mode effects and higher order mutual relationships through specialized projection operators. Numerical evaluations focusing on three distinct benchmarks from various fields demonstrate that the proposed HMD framework consistently yields significantly lower reconstruction errors compared to both Tucker and CP decomposition. These results establish HMD as a robust, high fidelity computational method for resolving complex, deeply coupled multidimensional data structures in science and engineering applications.
\end{abstract}

% Keywords
%\begin{IEEEkeywords}
%Holistic Multivariance Decomposition, Tensor Decomposition, Low-rank Approximation, Enhanced Multivariance Products Representation, Mode interrelations
%\end{IEEEkeywords}

% Main text
\section{Introduction}
\label{sec:intro}

Modern scientific and engineering applications generate vast amounts of multidimensional data, requiring a crucial need for advanced mathematical frameworks that can preserve inherent data geometries. Across disciplines ranging from signal processing \citep{signalTensor} to earth observation \citep{hsiTensor,tunaTGRS,tunaFranklin}, and biomedical engineering \citep{biomedicalTensor} to spectroscopy \citep{chemoTensor}, data is naturally captured across multiple coupled modes. In hyperspectral imaging, for example, scenes are recorded across two spatial dimensions and hundreds of contiguous spectral bands, forming massive, intrinsically $3$-D  structures \citep{hsiTensor}. Similarly, in fluorescence spectroscopy, tensors naturally emerge when analyzing chemical mixtures; here, the first dimension tracks the individual samples, while the second and third dimensions represent the emission and excitation wavelengths, respectively \citep{chemoTensor}. 

Historically, analytical techniques have relied on matricization (unfolding) to process these datasets. However, flattening multidimensional arrays into $2$-D matrices inevitably disrupts vital spatial-spectral correlations, leading to significant information loss and suboptimal pattern extraction \citep{tensor2d}. To prevent this, multiway arrays, or tensors, have emerged as the standard mathematical framework for representing and analyzing multidimensional data.

The central challenge in multilinear algebra lies in identifying and extracting latent patterns within a given tensor through computationally efficient decompositions. The foundations of this domain rest on the Tucker decomposition \citep{koldaTensor,signalTensor} and the Canonical Polyadic (CP) decomposition \citep{koldaTensor,signalTensor}. The CP model expresses a tensor as a minimal sum of rank-$1$ components \citep{koldaTensor,signalTensor}, while the Tucker model decomposes a tensor into a core tensor multiplied by orthogonal projection matrices along each mode, effectively isolating the principal subspaces of the multidimensional data \citep{koldaTensor,signalTensor}.

A substantial portion of modern multi-way data analysis focuses on low-rank approximation phenomena, which fundamentally aim to reconstruct a high dimensional tensor using a truncated summation of elementary components \citep{low_rankreview}. In classical CP and Tucker frameworks, these components are structurally established by computing the outer product of individual vectors drawn from each independent dimension (mode, way). While mathematically elegant and computationally tractable, this classical paradigm suffers from a severe structural limitation. Because the underlying terms are rigidly tied to pure multilinear outer products, each rank-$1$ component is only capable of capturing the isolated, independent variations of individual modes. Consequently, traditional low-rank approximations inherently overlook the complex, mutual relationships and highly coupled interdependencies that characterize real world physical systems \citep{low_rankreview}. When applied to complex datasets, this independence assumption forces the decomposition to sacrifice structural fidelity, failing to encapsulate the shared cross-mode dynamics that define multidimensional phenomena.

To address the limitations of assuming strictly decoupled modal behaviors, researchers have turned to multivariance analysis frameworks explicitly designed to map higher order interactions. Originating in function approximation, techniques such as High Dimensional Model Representation (HDMR) decompose complex input-output systems by systematically separating independent variable influences from their cooperative, multivariable interactions \citep{tunaHDMR, burcuHDMR}. Bridging this concept to discrete datasets, the Enhanced Multivariance Product Representation (EMPR) was developed to provide a more descriptive, non-local expansion of multi-way arrays (tensors). Unlike standard decompositions that rely purely on rigid basis matrices, EMPR integrates flexible $1$-D support vectors directly into its representation, allowing the framework to inherently scale from independent modal behaviors up to intricate, multidimensional cooperative terms \citep{tunaTGRS,tunaFranklin,tunaTCBB,tunaCNN}.

Recent literature demonstrates the exceptional efficacy of EMPR in solving highly complex, tensor focused engineering problems where traditional methods fail. For instance, the framework has been successfully deployed for the lossy compression of high dimensional hyperspectral images, where it achieves high compression ratios by explicitly and iteratively resolving intense spatial-spectral correlations \citep{tunaTGRS}. Beyond data reduction, EMPR has shown superior performance in robust feature extraction schemes for remote sensing data, utilizing strategic support optimization to isolate target signatures from dense background noise \citep{tunaFranklin}. Extending these capabilities to the genomics domain, researchers have also established efficient EMPR based feature selection schemes to untangle the highly correlated pathways within gene expression networks \citep{tunaTCBB}. Moreover, in deep learning architectures, EMPR has been utilized to compress the dense weight tensors of Convolutional Neural Networks (CNNs), drastically lowering structural redundancy to optimize both training and inference efficiency across various image analysis tasks \citep{tunaCNN}.

While EMPR offers vastly superior variance mapping attributes compared to baseline models, integrating these cooperative multivariance structures directly into a generalized, unified low-rank tensor decomposition framework remains a highly challenging computational and algebraic task. To bridge the theoretical gap between discrete multivariance representations and formal multilinear algebra, we introduce the Holistic Multivariance Decomposition (HMD) method in this work.

HMD introduces a novel, hierarchical approximation scheme that simultaneously accounts for both the isolated effects of individual modes and the mutual, synergistic relationships between dimensions. By formulating specialized projection operators, our framework isolates and retains the dimensional synergy that standard rank-$1$ outer product summations inherently omit. We demonstrate that, across a diverse range of multidimensional data characteristics spanning hyperspectral imaging, dynamic video sequences and chemometrics, HMD significantly outperforms traditional Tucker and CP decompositions in both reconstruction accuracy and structural quality. Notably, in existing multivariance literature, researchers frequently must resort to complex, computationally demanding second order EMPR approximants to outperform state-of-the-art (SoTA) methods \citep{tunaTGRS, tunaHDMR, sota_3dspeck}. In contrast, the mathematical elegance of the HMD framework enables it to decisively surpass SoTA tensor decomposition techniques while only taking its first level terms into account, representing a major leap forward in computational efficiency and structural expression.

The remainder of this paper is organized as follows. Section \ref{sec:bg} establishes the mathematical notation and briefly reviews foundational tensor operations alongside with the Tucker decomposition and EMPR. Section \ref{sec:HMD} details the theoretical derivation of the Holistic Multivariance Decomposition framework, which is followed in Section \ref{sec:experiments} by a rigorous mathematical analysis of its associated truncation and projection errors. We discuss our numerical experiments and benchmark the proposed technique against classical methods using relevant engineering datasets herein. Finally, Section \ref{sec:conc} concludes the paper and outlines future directions for research.

\section{Background}
\label{sec:bg}

\subsection{Tucker Decomposition}
\label{subsec:tucker}

Tucker decomposition is one of the foundational tensor decomposition techniques in multilinear algebra \citep{koldaTensor,koldaMultilinear,tensor2d}. It enables the representation of a given multiway array as a linear combination of rank-$1$ terms, which are constructed via the outer products of the columns of specific factor matrices. These factor matrices are typically tall (vertically oriented), meaning their number of columns is strictly and substantially smaller than their number of rows.

To be precise, let $\mathcal{X}$ denote a $D$-dimensional tensor of size $n_1 \times n_2 \times \dots \times n_D$, and let $\mathbf{A}^{(k)}$ represent the $k$-th factor matrix of size $n_k \times r_k$, such that
%---------------------------------------------
\begin{equation}
\mathcal{X} \in \mathbb{R}^{n_1 \times n_2 \times \dots \times n_D}, \quad \mathbf{A}^{(k)} \in \mathbb{R}^{n_k \times r_k}; \quad r_k < n_k, \quad k = 1, 2, \dots, D
\label{eq:tucker_setup}
\end{equation}
%---------------------------------------------
The Tucker decomposition of the tensor $\mathcal{X}$ is explicitly defined as follows
%---------------------------------------------
\begin{equation}
\mathcal{X} \approx \mathcal{G} \times_1 \mathbf{A}^{(1)} \times_2 \mathbf{A}^{(2)} \times_3 \dots \times_D \mathbf{A}^{(D)} = \sum_{i_1=1}^{r_1} \sum_{i_2=1}^{r_2} \dots \sum_{i_D=1}^{r_D} g_{i_1 i_2 \dots i_D}\, \mathbf{a}_{i_1}^{(1)}\, \circ\, \mathbf{a}_{i_2}^{(2)}\, \circ\, \dots\, \circ\, \mathbf{a}_{i_D}^{(D)}
\label{eq:tucker_dec}
\end{equation}
%---------------------------------------------
where $\mathcal{G}$ is the core tensor of size $r_1 \times r_2 \times \dots \times r_D$, and $\times_k$ denotes the mode-$k$ product operator along the $k$-th mode whose explicit definition is as follows
%---------------------------------------------
\begin{equation}
\left(\mathcal{X} \times_n \mathbf{A}\right)_{i_1,\ldots,i_{n-1}j\,i_{n+1},\ldots,i_N}
=
\sum_{i_n=1}^{I_n}
\mathcal{X}_{i_1 i_2 \cdots i_N}\, A_{j i_n}.
\label{eq:moden}
\end{equation}
%---------------------------------------------
Moreover, $\mathbf{a}_{i_k}^{(k)}$ stands as the $i_k$-th column vector of the $k$-th factor matrix $\mathbf{A}^{(k)}$. The choice of the ranks $r_k$ determines the overall rank order of the Tucker approximation for the given tensor $\mathcal{X}$.

\begin{figure}[htbp]
  \centering
    \includegraphics[scale=0.3]{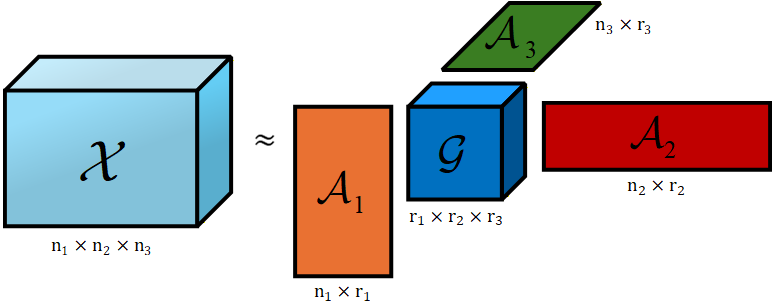}
    \caption{Tucker decomposition demonstration for $3$D tensors}\label{fig:tucker}
\end{figure}

A graphical illustration of the Tucker decomposition for a representative $3$ dimensional tensor $\mathcal{X}$ is displayed in Fig. \ref{fig:tucker} .
%---------------------------------------------
%---------------------------------------------

As can be observed from both Eqn.  \ref{eq:tucker_dec} and Fig. \ref{fig:tucker}, the classical approach initially isolates the effect of each mode before attempting to ensemble them via outer product and addition operations. Consequently, this architectural framework inherently fails to explicitly reflect the mutual interrelations among modes when reconstructing the underlying tensor structure.

\subsection{Enhanced Multivariance Products Representation}
\label{subsec:empr}

Enhanced Multivariance Products Representation (EMPR) is an efficient tensor decomposition technique that represents a given tensor in terms of lower dimensional entities such as %
scalers, vectors and matrices. By uniquely accounting for the cross-interrelations among modes, EMPR has found numerous applications across various fields ranging from data compression to feature extraction where high dimensionality is inevitable.

\begin{figure}[htbp]
  \centering
    \includegraphics[scale=0.3]{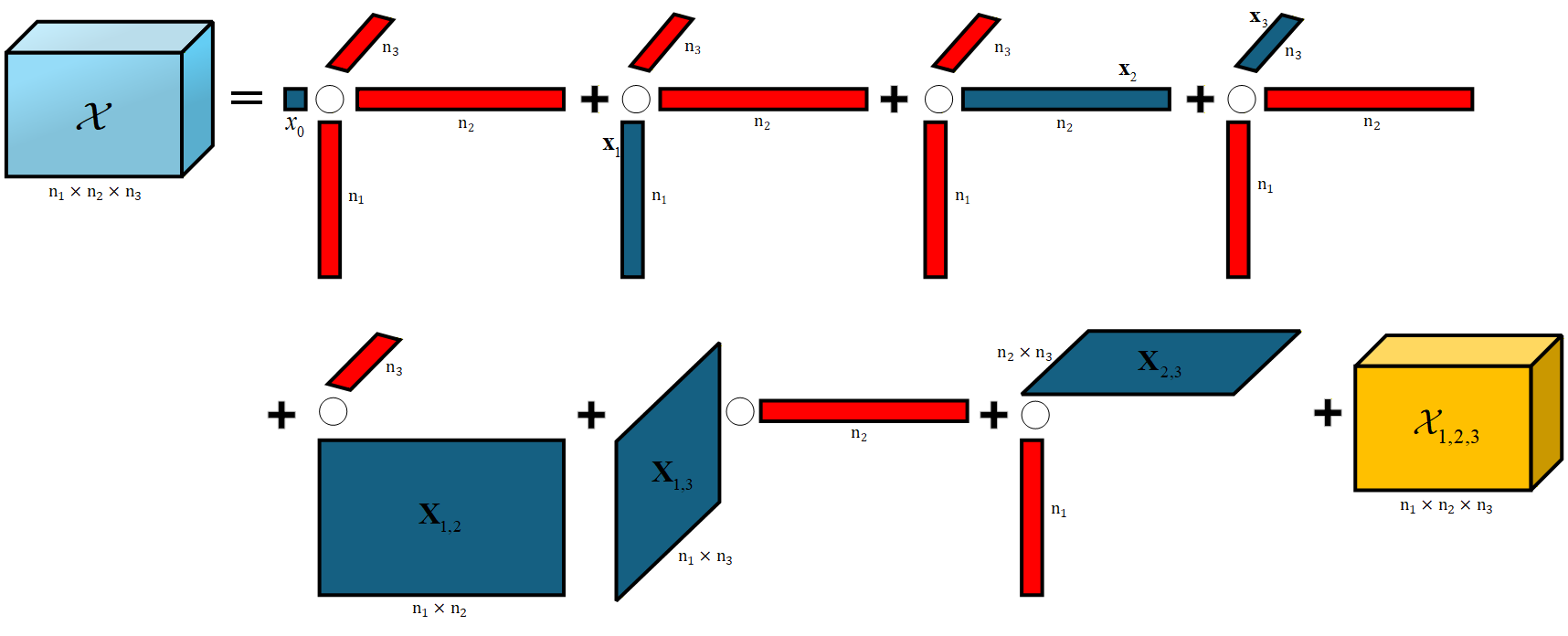}
    \caption{Enhanced Multivariance Products Representation demonstration for $3$D tensors}\label{fig:empr}
\end{figure}

The EMPR expansion of a given $D$-dimensional tensor $\mathcal{X}$ is explicitly defined as follows 
%---------------------------------------------
\begin{align}
\label{eq:empr}
\mathcal{X}_{i_1 \dots i_D} &= \mathcal{X}^{(0)}\left(\prod_{j=1}^D s_{i_j}^{(j)}\right)  + \sum_{j_1=1}^D \mathcal{X}_{i_{j_1}}^{(j_1)} \left(\prod_{\substack{j=1 \\ j \neq j_1}}^D s_{i_j}^{(j)}\right) + \sum_{\substack{j_1, j_2=1 \\ j_1 < j_2}}^D \mathcal{X}_{i_{j_1}, i_{j_2}}^{(j_1 j_2)}\left(\prod_{\substack{j=1 \\ j \neq j_1, j_2}}^D s_{i_j}^{(j)}\right)
\\ \nonumber &+ \dots + \mathcal{X}_{i_{j_1} i_{j_2} \dots i_{j_D}}^{(j_1 j_2 \dots j_D)}
\qquad i_j = 1, 2, \dots, n_j, \quad j = 1, 2, \dots, D
\end{align}
%---------------------------------------------
where the subscripts denote indices and the superscripts within parantheses indicate the 
modes to which the corresponding entity relates. In Eqn. \ref{eq:empr}, $s^{(j)}$'s are 
called support vectors since they depend on a single index while $\mathcal{X}^{(0)}$, %
$\mathcal{X}^{(j_1)}$, $\mathcal{X}^{(j_1,j_2)}$, $\ldots$ are designated as EMPR components. The graphical illustration of the EMPR expansion for a $3$-D tensor $\mathcal{X}$ is 
presented in Fig. \ref{fig:empr}  
%---------------------------------------------
%---------------------------------------------
where the blue elements depict the EMPR components while the red vectors represent the 
support vectors. As shown in Fig. \ref{fig:empr}, $x_0$ is the scalar EMPR component; $x_1$, $x_2$ and $x_3$ are the vector EMPR components while $X_{1,2}$, $X_{1,3}$ and $X_{2,3}$ constitute the matrix EMPR components. The final yellow term $\mathcal{X}_{1,2,3}$ is the residual term which shares the same size and dimensions as the original tensor to be decomposed $\mathcal{X}$. 
In practical applications, the residual term is typically truncated because it is highly 
susceptible to high dimensionality, correlation and noise \citep{tunaTGRS, tunaFranklin, burcuEMPR}.

To regulate the contribution of each entity within the EMPR framework, normalized weight vectors along each dimension are applied such as 
%---------------------------------------------
\begin{equation}
\label{eq:weights}
\sum_{i_j=1}^{n_j} W_{i_j}^{(j)} = 1, \qquad W_{i_1 \dots i_N} \equiv \prod_{j=1}^D W_{i_j}^{(j)}; \qquad j = 1, 2, \dots, D
\end{equation}
%---------------------------------------------
where an overall purely multiplicative weight tensor is established as described in Eqn. \ref{eq:weights}.

Support vectors are critical to EMPR decomposition since they directly affect the quality of the resulting representation. These preselected one-dimensional elements are integrated into the decomposition framework, where each support vector must satisfy the following normalization condition under the weight vector of its respective mode
%---------------------------------------------
\begin{equation}
\label{eq:supnorm}
\sum_{i_j=1}^{n_j} W_{i_j}^{(j)} \left(s_{i_j}^{(j)}\right)^2 = 1; \qquad j = 1, 2, \dots, D
\end{equation}
%---------------------------------------------
To ensure unique EMPR expansion terms for a given set of weight vectors and preselected support vectors, each term must satisfy the following vanishing conditions
%---------------------------------------------
\begin{equation}
\label{eq:vanishing}
\sum_{i_{j_\ell}=1}^{n_{j_\ell}} W_{i_{j_\ell}}^{(j_\ell)} s_{i_{j_\ell}}^{(j_\ell)} \mathcal{X}_{i_{j_1} \dots i_{j_k}}^{(i_{j_1} \dots i_{j_k})} = 0; \qquad \ell = 1, 2, \dots, k, \quad k = 1, 2, \dots, D
\end{equation}
%---------------------------------------------
By virtue of the conditions in Eqn. \ref{eq:vanishing}, the EMPR terms become mutually %
orthogonal under the specified weight vectors \citep{tunaTGRS,tunaTCBB}. Consequently, by combining the conditions from Eqn. \ref{eq:weights}, Eqn. \ref{eq:supnorm} and Eqn. \ref{eq:vanishing}, the scalar EMPR term $\mathcal{X}_0$ can be calculated as 
%---------------------------------------------
\begin{equation}
\label{eq:empr0}
\mathcal{X}^{(0)}=\sum_{j_1=1}^{n_1} \dots \sum_{j_D=1}^{n_D}
\left(\prod_{\ell=1}^D W_{i_{j_\ell}}^{(j_\ell)} s_{i_{j_\ell}}^{(j_\ell)} \right)
\mathcal{X}_{i_{j_1} \dots i_{j_D}}^{(i_{j_1} \dots i_{j_D})}
\end{equation}
%---------------------------------------------
Evidently, $\mathcal{X}_0$ represents a weighted average of the entries of tensor $\mathcal{X}$. As a result, $\mathcal{X}_0$ is typically treated as either a bias term or 
the coarsest approximation of tensor $\mathcal{X}$. 

Following the calculation of the scalar term, the one-way (vector) EMPR terms are determined by taking the weighted average of the entries of $\mathcal{X}$ accross alldimensions except the $k$-th one and subtracting the influence of the $k$-th support vector scaled by the 
constant scalar term as follows
%---------------------------------------------
\begin{align}
\label{eq:empr1}
\mathcal{X}_{i_k}^{(k)} = \sum_{j_1=1}^{n_1} \dots \sum_{j_{k-1}=1}^{n_{k-1}}
&\sum_{j_{k+1}=1}^{n_{k+1}} \dots \sum_{j_D=1}^{n_D} 
\left(\prod_{\substack{\ell=1 \\ \ell \neq k}}^D W_{j_\ell}^{(\ell)}s_{i_\ell}^{(\ell)}\right) \mathcal{X}_{j_1 \dots j_D} - \mathcal{X}^{(0)}\,s_{i_k}^{(k)};\nonumber \\
&i_k = 1, 2, \dots, n_k, \quad k = 1, 2, \dots, D
\end{align}
%---------------------------------------------

The $k$-th one-way EMPR term reflects the attitude of the multidimensional tensor 
$\mathcal{X}$ along the $k$-th dimension isolated from the rest of the all dimensions. Therefore, the one-way terms capture the individual model behaviour exclusive of joint interactions. 

Similarly, the two-way (matrix) terms isolate the specific interactions between pairs of distinct modes, while higher dimensional EMPR terms capture interactions across multiple modes. This capability uniquely distinguishes EMPR from the methods like Tucker and CP decompositions, which do not inherently capture the cross-interrelational behavior across all dimensions of a given tensor. Following a strategy similar to the one outlined in Eqn. \ref{eq:empr0} and Eqn. \ref{eq:empr1}, the two-way and higher dimensional EMPR terms can be uniquely computed under the aforementioned conditions.

As previously noted, the selection of support vectors is crucial for establishing highly effective EMPR representations. To address this, Averaged Directional Supports (ADS) are widely used in various applications \citep{tunaTGRS, tunaTCBB}. Alternatively, several methods for determining optimal supports have been explored in the literature \citep{tunaFranklin}. The ADS approach computes the weighted average of a tensor along a target mode, with its explicit definition formulated as follows in
%---------------------------------------------
\begin{equation}
s_{i_j}^{(j)} = \frac{\displaystyle\sum_{i_1=1}^{n_1} \dots \sum_{i_{j-1}=1}^{n_{j-1}} \sum_{i_{j+1}=1}^{n_{j+1}} \dots \sum_{i_D=1}^{n_D} 
\left(\prod^{n_\ell}_{\substack{\ell=1 \\ \ell\neq j}} W_{i_\ell}^{(\ell)} \right) \mathcal{X}_{i_1, \dots, i_D}}
{\displaystyle\sum_{i_j=1}^{n_j}\left( W_{i_j}^{(j)} \left[ \sum_{i_1=1}^{n_1} \dots \sum_{i_{j-1}=1}^{n_{j-1}} \sum_{i_{j+1}=1}^{n_{j+1}} \dots \sum_{i_D=1}^{n_D} 
\left(\prod^{n_\ell}_{\substack{\ell=1 \\ \ell\neq j}} W_{i_\ell}^{(\ell)} \right)
 \mathcal{X}_{i_1, \dots, i_D} \right]^2 \right)^{\frac{1}{2}}};\quad j=1,\ldots,D
\end{equation}
%---------------------------------------------
where each support vector $s^{(j)}$ is properly normalized to comply with Eqn. \ref{eq:supnorm}.

\section{Holistic Multivariance Decomposition}
\label{sec:HMD}

Classical low-rank approximation techniques based on the Tucker and CP methods are not well suited to handle cross-interrelational behaviors among tensor modes. Conversely, while EMPR can account for these behaviors, it possesses limited representation capability because the rank of its expansion is unadjustable due to the one-dimensional structure of the support vectors defined in \ref{eq:supnorm}. To combine the interrelational, intrinsic feature extraction capability of EMPR with the rank adjustment flexibility of the Tucker decomposition, we propose a novel and efficient tensor decomposition method named Holistic Multivariance Decomposition (HMD).  
 
Let $\mathcal{X} \in \mathbb{R}^{n_1 \times \dots \times n_D}$ is a $D$-dimensional  
tensor and $S_i$ be matrices such that 
%---------------------------------------------
\begin{equation}
S_i \in \mathbb{R}^{n_i \times r_i}\,; \qquad r_i < n_i\,, \qquad i=1, \dots, D
\label{eq:Si}
\end{equation}
%---------------------------------------------
where each $S_i$ satisfies the following condition,
%---------------------------------------------
\begin{equation}
\label{eq:Sorth}
S_i^T\, S_i = n_i\,I_{r_i \times r_i}; \qquad i=1, \dots, D
\end{equation}
%---------------------------------------------
Evidently, each $S_i$ is a scaled semi-orthogonal vertically rectangular matrix.  

To simplify the notation, let $N$ denote the product of the sizes of the tensor $\mathcal{X}$. 
We define $N^{(i)}$ as the product of all mode sizes except the $i$-th dimension, 
and $N^{(i,j)}$ as the product of all mode sizes excluding both the $i$-th and 
$j$-th dimensions. Their explicit definitions can be written as follows
%---------------------------------------------
\begin{equation}
\label{eq:Ns}
N = \prod_{k=1}^D n_k,\qquad
N^{(i)} = \prod_{\substack{k=1 \\ k \neq i}}^D n_k,\qquad
N^{(i,j)} = \prod_{\substack{k=1 \\ k \neq i,j}}^D n_k,\qquad\cdots
\end{equation}
%---------------------------------------------

Let $\mathcal{S}$ represent the set of matrices described in Eqn. \ref{eq:Si} and 
\ref{eq:Sorth} as
%---------------------------------------------
\begin{equation}
\mathcal{S} = \left\{S_1, \dots, S_D\right\}
\end{equation}
%---------------------------------------------
Following the same logic as in Eqn. \ref{eq:Ns}, the subset of $\mathcal{S}$ 
that excludes the $i$-th matrix, or both the $i$-th, $j$-th matrices are defined as 
%---------------------------------------------
\begin{equation}
\label{eq:Sis}
\mathcal{S}^{(i)} = \left\{S_1, \dots, S_{i-1}, S_{i+1}, \dots, S_D\right\}
\end{equation}
%---------------------------------------------
and
%---------------------------------------------
\begin{equation}
\label{eq:Sijs}
\mathcal{S}^{(i,j)} = \left\{S_1, \dots, S_{i-1}, S_{i+1}, \dots, S_{j-1}, S_{j+1}, \dots, S_D\right\}
\end{equation}
%---------------------------------------------
respectively.

Consider the following definition,
%---------------------------------------------
\begin{equation}
\label{eq:tuckerop}
\kruskal{\mathcal{X}; \mathcal{S}} = \kruskal{\mathcal{X}\,;\,\left\{S_1, \dots, S_D\right\} } = \mathcal{X} \times_1 S_1 \times_2 \dots \times_D S_D
\end{equation}
%---------------------------------------------
where $\mathcal{X}$ is a $D$-dimensional tensor and $S_i$'s are the matrices that %
enable mode-$n$ product of tensor $\mathcal{X}$ with the corresponding $S_i$ matrix. The notation 
$\kruskal{\bullet}$ is widely referred as the Tucker operator in the literature \citep{koldaMultilinear}.

Utilizing the Tucker operator, the Holistic Multivariance Decomposition of a given tensor %
$\mathcal{X}$ is defined as follows
%---------------------------------------------
\begin{equation}
\label{eq:hmd}
\mathcal{X} = \kruskal{\mathcal{X}_0\,;\, \mathcal{S} } + \sum_{i=1}^D \kruskal{ \mathcal{X}_i\,;\, \mathcal{S}^{(i)} } 
 + \sum_{\substack{i,j=1 \\ i < j}}^D \kruskal{ \mathcal{X}_{i,j}\,;\, \mathcal{S}^{(i,j)} } + \dots
\end{equation}
%--------------------------------------------- 
In Eqn. \ref{eq:hmd}, $\mathcal{X}_0$ is denoted as the zeroth level HMD component while %
$\mathcal{X}_i$'s are the first level HMD components. $\mathcal{X}_{i,j}$'s represent %
the second level HMD components and all higher level components can be named in 
an analogous fashion. 

\begin{figure}[htbp]
  \centering
    \includegraphics[scale=0.28]{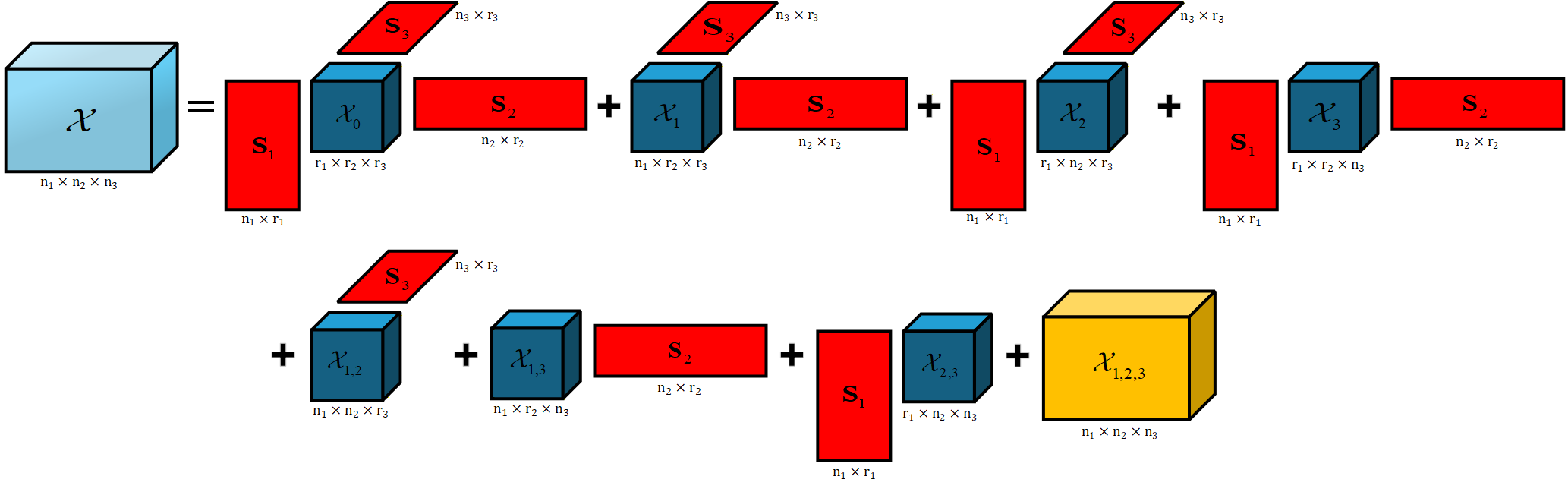}
    \caption{Holistic Multivariance Decomposition demonstration for $3$D tensors}\label{fig:hmd}
\end{figure}

In contrast to EMPR, the HMD components do not consist of scalars, 
vectors, matrices and so on. Instead, they are all $D$-dimensional tensors contain a relatively  small number of entries due to condition in \ref{eq:Si}. A $3$-D illustration of the HMD expansion is displayed in Fig. \ref{fig:hmd}, 
%---------------------------------------------
%---------------------------------------------
where the blue blocks represent the HMD components and the red entities denote %
the support matrices. The dimensions of each entity can be observed in Fig. \ref{fig:hmd}. 
Since the number of columns in each support matrix is strictly less than its number 
of rows, each HMD term is a tensor whose total number of entries is significantly lower than the original tensor $\mathcal{X}$. Consequently, the zeroth level tensor $\mathcal{X}_0$in Fig. \ref{fig:hmd} has dimensions of $r_1\times r_2\times r_3$, whereas the first level HMD terms $\mathcal{X}_1$, $\mathcal{X}_2$ and $\mathcal{X}_3$ have sizes of $n_1\times r_2\times r_3$, $r_1\times n_2\times r_3$ and $r_1\times r_2\times n_3$, respectively.

In Subsection \ref{subsec:empr}, we discussed the use of weight to adjust %
contributions along each tensor mode. A similar approach can be adapted for HMD. 
To simplify the subsequent analysis, we fix the weight elements across each mode 
to be invariant, setting $W^{(i)}=\frac{1}{n_i}$ for $i=1,\ldots,D$. 

Tensors and matrices obey the following properties with respect to the mode-$n$ product \citep{koldaTensor}. The first is the commutativity property and expressed as 
%---------------------------------------------
\begin{equation}
\label{eq:modenm}
\mathcal{X}\times_m A\times_n B = \mathcal{X}\times_n B\times_m A,\qquad m\neq n
\end{equation}
%---------------------------------------------
while the second identity (benefiting from Eqn. \ref{eq:Sorth}) yields
%---------------------------------------------
\begin{equation}
\label{eq:modenn}
\mathcal{X}\times_n S\times_n S = \mathcal{X}\times_n \left(S^T\,S\right)
\end{equation}
%---------------------------------------------.

If we apply the Tucker operator from Eqn. \ref{eq:tuckerop} using the complete set of support matrices $\mathcal{S}$ to both sides of the expansion in 
Eqn. \ref{eq:hmd}, and multiply by all constant $W^{(i)}$ weights, the zeroth level HMD term 
$\mathcal{X}_0$ can be computed as 
%---------------------------------------------
\begin{equation}
\label{eq:X0}
\mathcal{X}_0 = \frac{1}{N} \kruskal{ \mathcal{X}\,;\, \mathcal{S} }
\end{equation}
%---------------------------------------------
The zeroth level HMD term serves as the direct multilinear analog to the scalar EMPR term $x_0$. However, their geometric treatments of the data are fundamentally different. While EMPR strictly collapses the entire multidimensional block into a single, dimensionless point, HMD compresses the tensor space into a dense core tensor of size $r_1 \times r_2 \dots \times r_D$. This approach ensures that the global multidimensional architecture of the data is safely preserved within a downsampled geometric subspace, rather than being completely flattened.

Following the same strategy, the first level HMD terms are obtained as 
%---------------------------------------------
\begin{equation}
\label{eq:Xi}
\mathcal{X}_i = \frac{1}{N^{(i)}} \kruskal{ \mathcal{X}\,;\, \mathcal{S}^{(i)} } - 
\kruskal{\mathcal{X}_0\,;\, S_i}; \quad i=1,\dots,D
\end{equation}
%---------------------------------------------
while the second level HMD terms are determined as follows
%---------------------------------------------
\begin{equation}
\label{eq:Xij}
\mathcal{X}_{i,j} = \frac{1}{N^{(i,j)}} \kruskal{ \mathcal{X}\,;\, \mathcal{S}^{(i,j)} } -
\kruskal{ \mathcal{X}_0\,;\, S_i, S_j} 
 - \kruskal{\mathcal{X}_i\,;\, S_j} - 
\kruskal{\mathcal{X}_j\,;\, S_i}; \qquad 
i,j=1,\dots,D, \qquad i < j
\end{equation}
%---------------------------------------------
Higher level HMD terms can be obtained in a similar fashion. 

The first level HMD terms defined in \ref{eq:Xi} conceptually mirror the one-way vector components of EMPR. Geometrically, EMPR reduces variations along a single mode into rigid, one-dimensional lines. In contrast, \ref{eq:Xi} executes a partial projection and deflation to isolate a full $D$-dimensional tensor subspace. Within this specific subspace, the informational rank is exclusively dictated by linear deviations along the $i$-th axis. It behaves geometrically much like a bundle of parallel fibers oriented along a single dimension, maintaining the full tensor structure instead of degrading it to a simple vector. 

The second level HMD terms in \ref{eq:Xij} expand upon the concept of the two-way EMPR matrix components. Where EMPR forces pairwise relationships to flatten into standard two-dimensional matrices, HMD preserves the geometry as an adjustable multidimensional tensor block. By systematically deflating the core and first level terms, this equation geometrically isolates a relational manifold embedded directly within the higher-dimensional tensor architecture.

By defining the index set $\mathbb{N}_D=\{1, \dots, D\}$ and considering all of its %
all subsets, the HMD expansion in Eqn. \ref{eq:hmd} can be reexpressed in a more elegant 
and compact form
%---------------------------------------------
\begin{equation}
\label{eq:HMDcompact}
\mathcal{X} = \bigsum_{\mathcal{I}\, \subseteq\, \mathbb{N}_D} \kruskal{ \mathcal{X}_\mathcal{I}\,;\, \mathcal{S}^{(\mathcal{I})} }; \qquad \mathbb{N}_D = \{1, \dots, D\}
\end{equation}
%---------------------------------------------
where $\mathcal{I}$ is a subset of $\mathbb{N}_D$ and the empty set corresponds to the 
zeroth level HMD term $\mathcal{X}_0$.

In practice, computing the complete expansion formulated in \ref{eq:HMDcompact} is computationally demanding and often unnecessary for high dimensional data. By systematically truncating the expansion and omitting higher level interaction terms, we obtain a highly efficient approximation. This approach retains the essential multivariance features of the tensor while completely avoiding combinatorial complexity.

\section{Experimental Results}
\label{sec:experiments}

To evaluate the efficiency of the proposed HMD method, we employ three $3$-D tensor datasets spanning distinct scientific fields. The first dataset is obtained from Hyperspectral Imagery, the second comprises video analysis, and the final dataset is drawn from chemometrics.

Furthermore, we benchmark our method against Tucker and CP decompositions to highlight the performance gains of HMD. As previously noted, it is computationally impractical to include all terms of the full expansion in \ref{eq:HMDcompact}. Therefore, we systematically truncate the expansion at specific structural levels. Initially, we consider only the zeroth level HMD term to establish a baseline zeroth level approximation. We then incorporate the three first level HMD terms to form the first level HMD approximation. Finally, adding the three second level terms yields the second level HMD approximant. To provide a comprehensive comparative analysis, all three of these sequential approximation levels are explicitly plotted.

For the numerical experiments, the support matrices $S_i$'s are constructed by extracting the dominant right singular vectors from the corresponding mode-$n$ unfolding of the target tensor. To ensure compliance with the essential semi-orthogonality requirements outlined in Eq. (\ref{eq:Sorth}), each resulting matrix is subsequently scaled by a factor of $\sqrt{n_i}$.

The first validation case utilizes the Indian Pines hyperspectral image dataset, which was captured via the NASA AVIRIS sensor \citep{indianData}. The original Indian Pines dataset consists of $220$ spectral bands. In this study, $20$ water absorption and low SNR bands were removed, resulting in a total of $200$ spectral bands used for experiments. Therefore, the tensor dimensions are $145 \times 145 \times 200$, with each spectral value stored at a $16$-bit precision. In this structure, the first two modes represent the spatial domain along the $x$ and $y$ axes, respectively, while the third mode contains the continuous spectral information.

%------------------Fig4------------------------
\begin{figure}[h!]
  \centering
    \includegraphics[scale=.5]{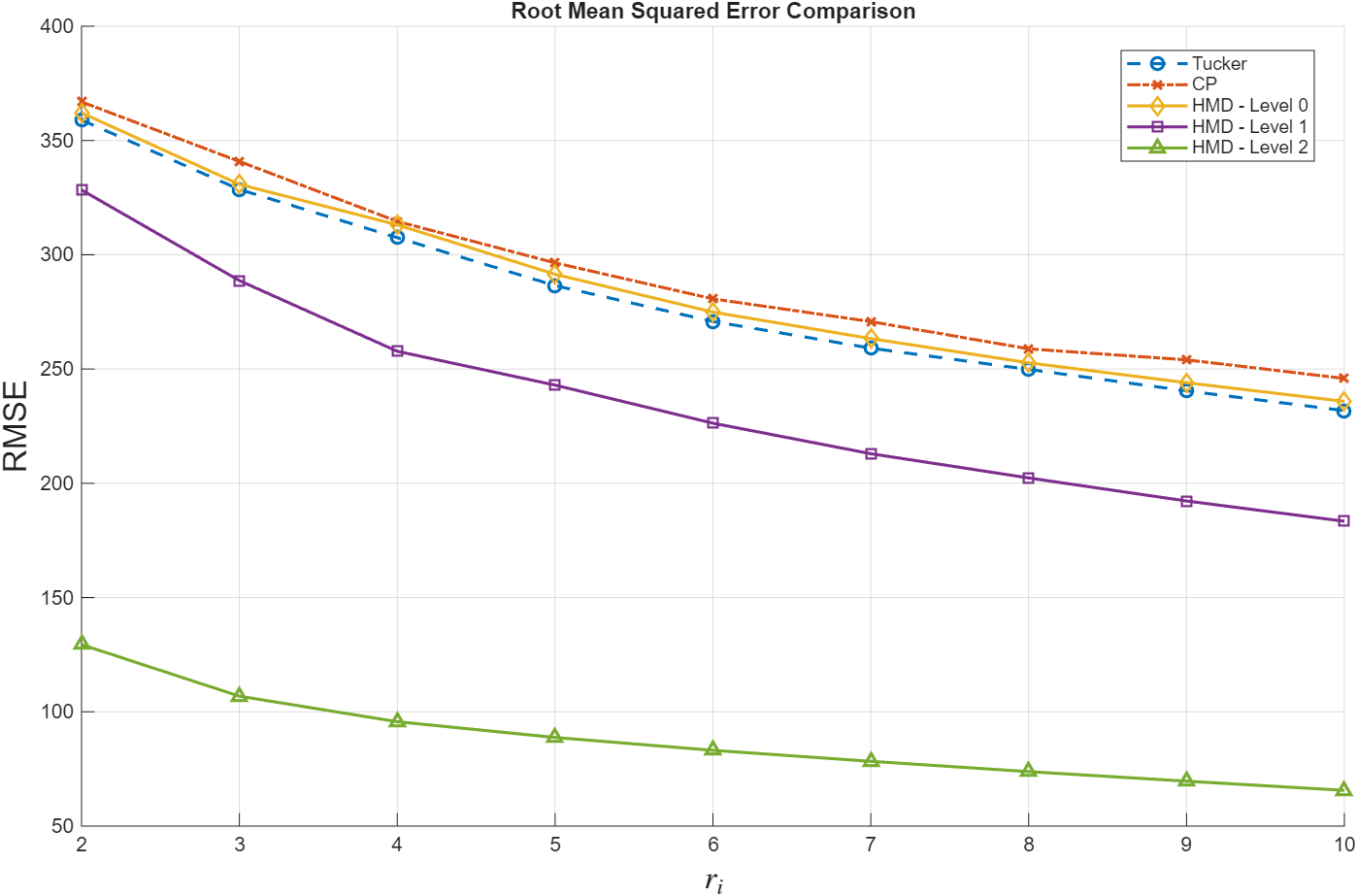}
    \caption{RMSE comparison as a function of rank $r_i$ for the Indian Pines hyperspectral dataset}
    \label{fig:IP_RMSE}
\end{figure}
%------------------Fig4------------------------

The capacity of the HMD method to preserve the structural fidelity of the original data is evaluated by analyzing the Root Mean Squared Error (RMSE) between the target dataset and its corresponding approximant. The RMSE metrics are calculated for the Tucker and CP methods, as well as for our sequential zeroth, first, and second level HMD expansions. For all tested methods, the rank values $r_i$ are varied from $2$ to $10$. Although the HMD framework incorporates three distinct support matrices each governed by its own independent rank parameter ($r_1$, $r_2$, and $r_3$), we fix these ranks to an identical value across all trials for a fair comparison.

As demonstrated in Fig. \ref{fig:IP_RMSE}, the resulting RMSE values decrease consistently as the rank $r_i$ increases. The CP method exhibits the highest reconstruction error, while the Tucker decomposition yields a marginal improvement over CP. Interestingly, even the highly compressed zeroth level HMD approximation performs competitively, tracking between the CP and Tucker baselines. More notably, the first level HMD approximant clearly outperforms both classical Tucker and CP decompositions. While the performance gap between the first level HMD and Tucker methods remains narrow at lower ranks, it widens significantly as $r_i$ grows. Specifically, at $r_i = 10$, the first level HMD achieves a reduced RMSE of $183.39$, whereas the Tucker decomposition yields a higher error value of $231.77$. Beyond these initial observations, a striking performance gap emerges when evaluating the second level HMD against the alternative frameworks. The numerical results indicate that even at a highly compressed rank of $r_i = 2$, the second level HMD achieves a significantly lower reconstruction error than both the Tucker and CP decompositions managed at their maximum tested rank of $r_i = 10$. Consequently, the trends illustrated in Fig. \ref{fig:IP_RMSE} emphasize that by incorporating pairwise mode interactions, the second level HMD provides a substantial performance boost, vastly outperforming traditional tensor decomposition paradigms.

Given that the Indian Pines dataset originates from the signal processing domain, it is highly instructive to investigate how effectively the proposed framework preserves signal quality and manages reconstruction distortion. To this end, we analyze the Peak Signal-to-Noise Ratio (PSNR) metrics across the different methods. 
%------------------Fig5------------------------
\begin{figure}[h!]
  \centering
    \includegraphics[scale=.5]{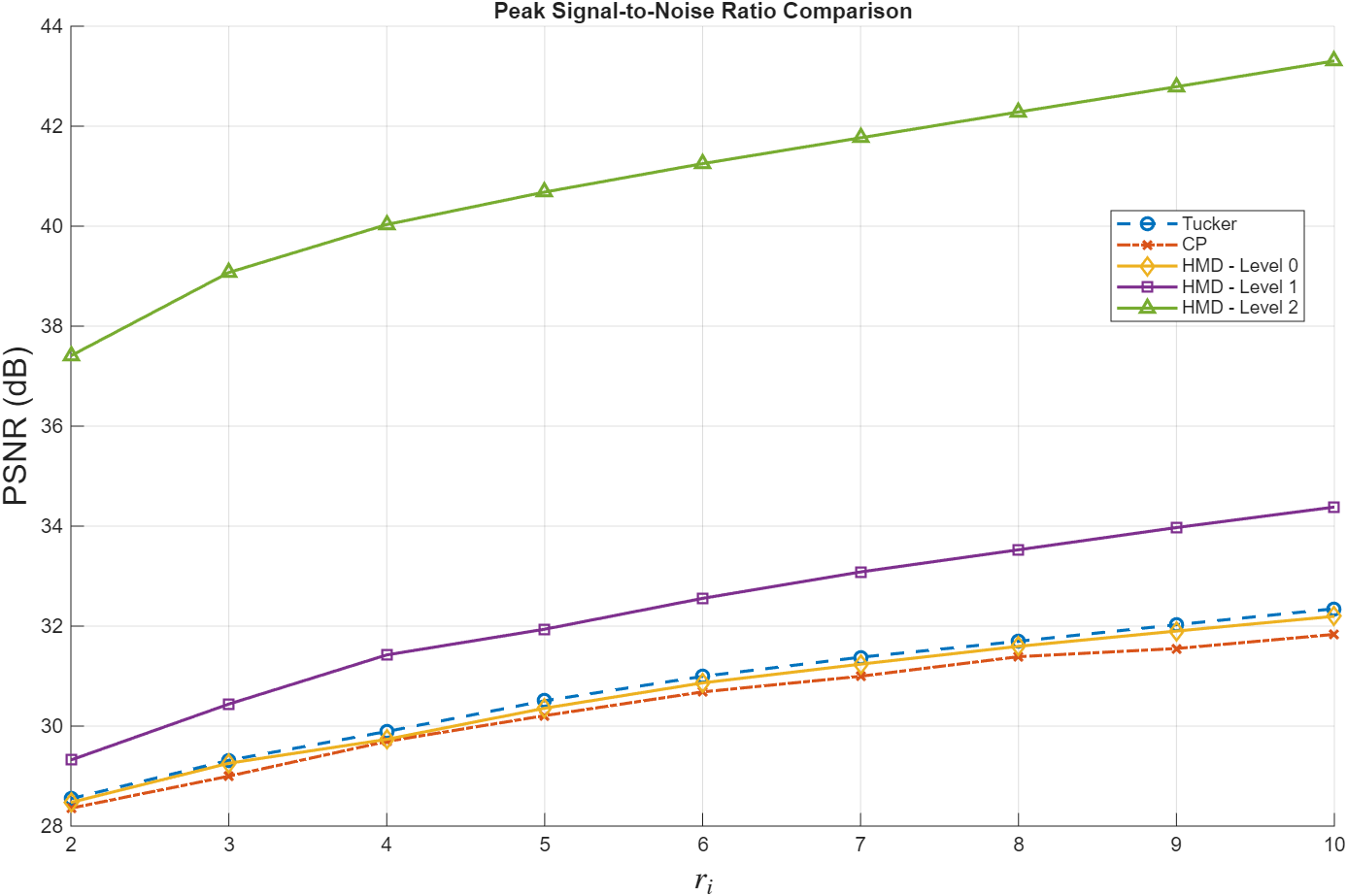}
    \caption{PSNR (dB) performance across varying ranks $r_i$ for the Indian Pines hyperspectral dataset }
    \label{fig:IP_PSNR}
\end{figure}
%------------------Fig5------------------------

As illustrated in Fig. \ref{fig:IP_PSNR}, the resulting trends closely mirror those observed in the RMSE analysis. The zeroth level HMD consistently tracks between the CP and Tucker decomposition baselines. Moving up the hierarchy, the first level HMD clearly outperforms both classical approaches, yielding higher PSNR values across all tested rank conditions. Most notably, the second level HMD delivers an exceptional performance leap relative to the other methods. It achieves a remarkable signal fidelity of $37.41$ dB at a very low rank of $r_i = 2$, and further boosts this quality to $43.31$ dB as the rank reaches $r_i = 10$.

Beyond their purely signal-oriented nature, hyperspectral datasets fundamentally preserve spatial image structures. Consequently, assessing the structural fidelity between the original image data and its reconstructed counterpart is crucial. This objective is effectively met by evaluating the Structural Similarity Index Measure (SSIM), a metric bounded between $0$ and $1$. An SSIM value approaching $1$ signifies that the reconstructed image is structurally nearly identical to the reference image, directly reflecting the superior performance of the underlying tensor decomposition framework.
%------------------Fig6------------------------
\begin{figure}[h!]
  \centering
    \includegraphics[scale=.5]{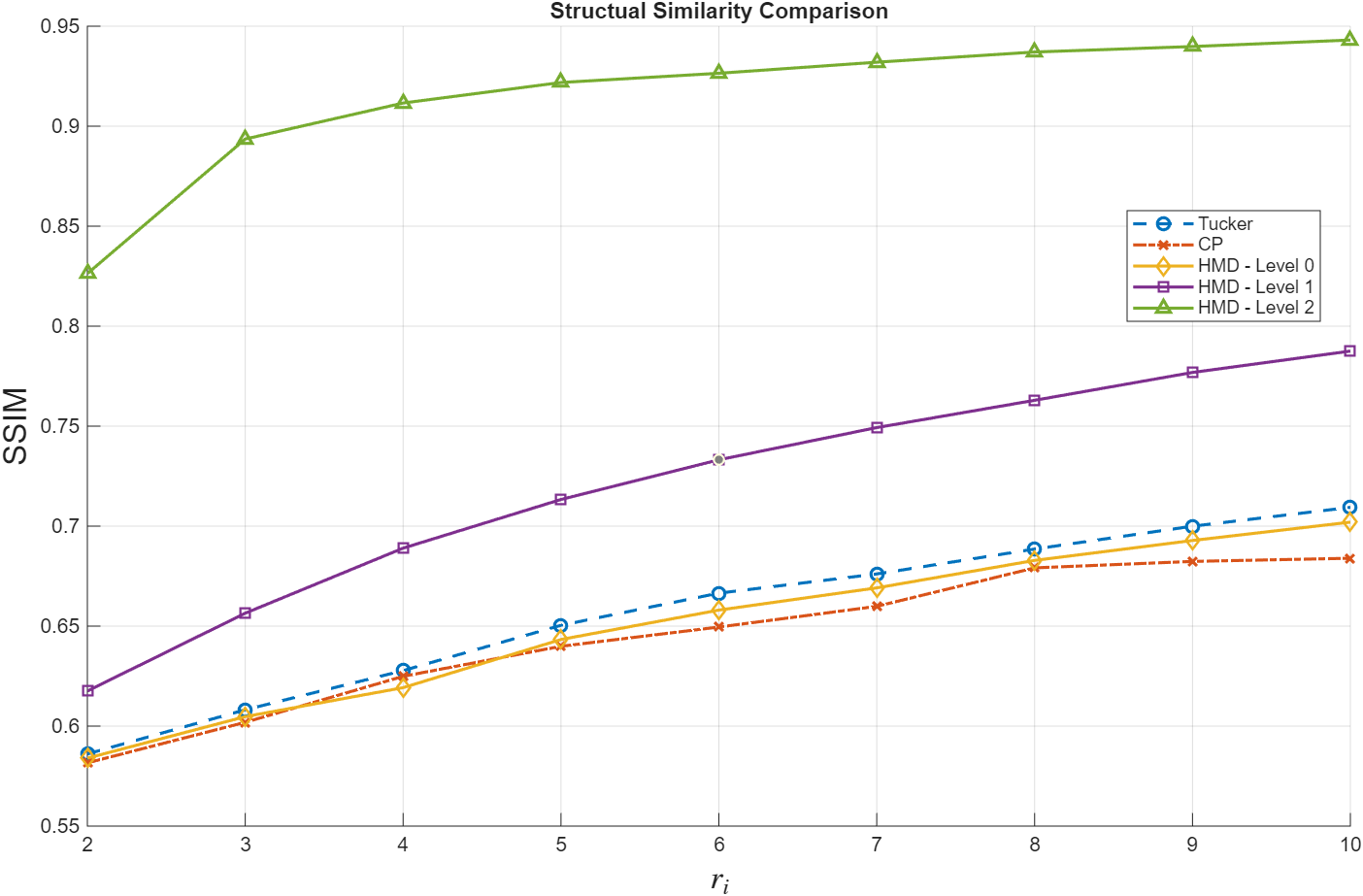}
    \caption{SSIM structural evaluation under different rank configurations $r_i$ for the Indian Pines hyperspectral dataset}
    \label{fig:IP_SSIM}
\end{figure}
%------------------Fig6------------------------
According to the metrics presented in Fig. \ref{fig:IP_SSIM}, the zeroth level HMD consistently tracks between the classical Tucker and CP baselines. Moving up the hierarchy, the first level HMD clearly outperforms both traditional methods, it achieves an SSIM value of $0.71$ at a modest rank of $r_i = 4$, whereas the Tucker decomposition requires its maximum tested rank of $r_i = 10$ to match this performance. Echoing the trends observed in the RMSE and PSNR analyses in Fig. \ref{fig:IP_RMSE} and Fig. \ref{fig:IP_PSNR}, respectively, Fig. \ref{fig:IP_SSIM} vividly highlights the exceptional performance leap delivered by the second level HMD. Initiating with a remarkably high structural similarity of $0.83$ at a minimal rank of $r_i = 2$, it increases steadily to reach a commanding $0.94$ at $r_i = 10$.

The second experiment utilizes the running action sequences from the publicly available KTH human action database developed by Schuldt, Laptev, and Caputo \citep{videoData}. The dataset comprises video recordings of $25$ subjects performing the specified action across four distinct environmental scenarios. All sequences were captured using a static camera at a frame rate of $25$ fps against homogeneous backgrounds, and subsequently downsampled to a spatial resolution of $160 \times 120$ pixels. For the specific sample under consideration, the video duration spans approximately $13$ seconds, yielding a $3$-D data tensor of size $160 \times 120 \times 345$ encoded at an $8$-bit precision.

%------------------Fig7------------------------
\begin{figure}[h!]
  \centering
    \includegraphics[scale=.5]{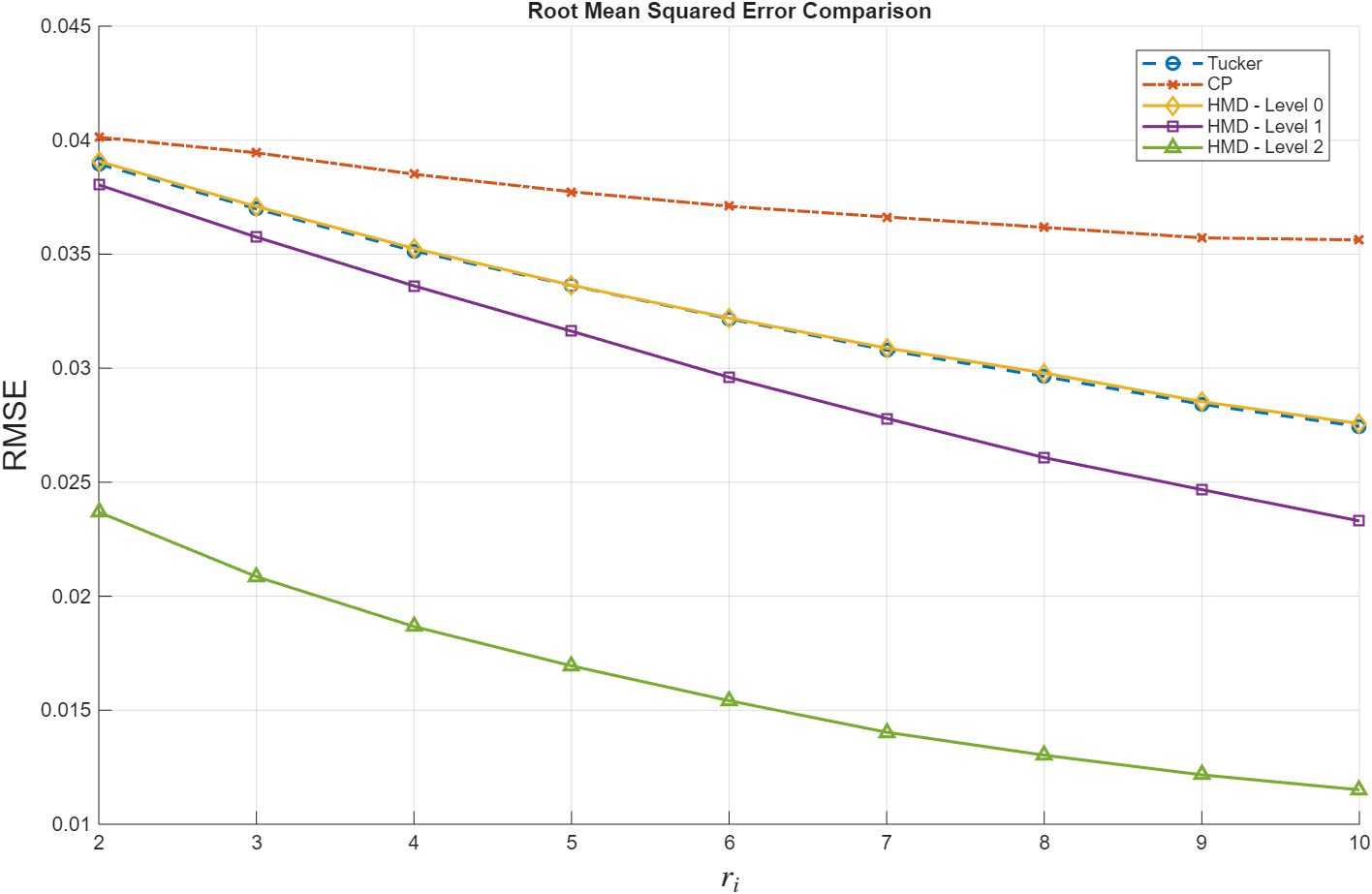}
    \caption{RMSE comparison as a function of rank $r_i$ for the KTH video dataset }
    \label{fig:Video_RMSE}
\end{figure}
%------------------Fig7------------------------

The RMSE evaluations in Fig. \ref{fig:Video_RMSE} illustrate the clear advantage of the HMD framework when processing dynamic video data, where spatial geometry is intricately coupled with temporal motion. While classical CP and Tucker decompositions struggle to compress this multimode variance, our hierarchical approach systematically lowers the approximation error. The zeroth and first level HMDs establish solid competitive baselines, but the second level HMD delivers a profound performance leap. By explicitly capturing the synergistic pairwise interactions between the spatial coordinates and the moving temporal frames, it preserves the underlying mechanics of the action sequence. This allows the second-level HMD to minimize residual error drastically, achieving exceptional accuracy even under aggressive low-rank conditions.

%------------------Fig8------------------------
\begin{figure}[h!]
  \centering
    \includegraphics[scale=.5]{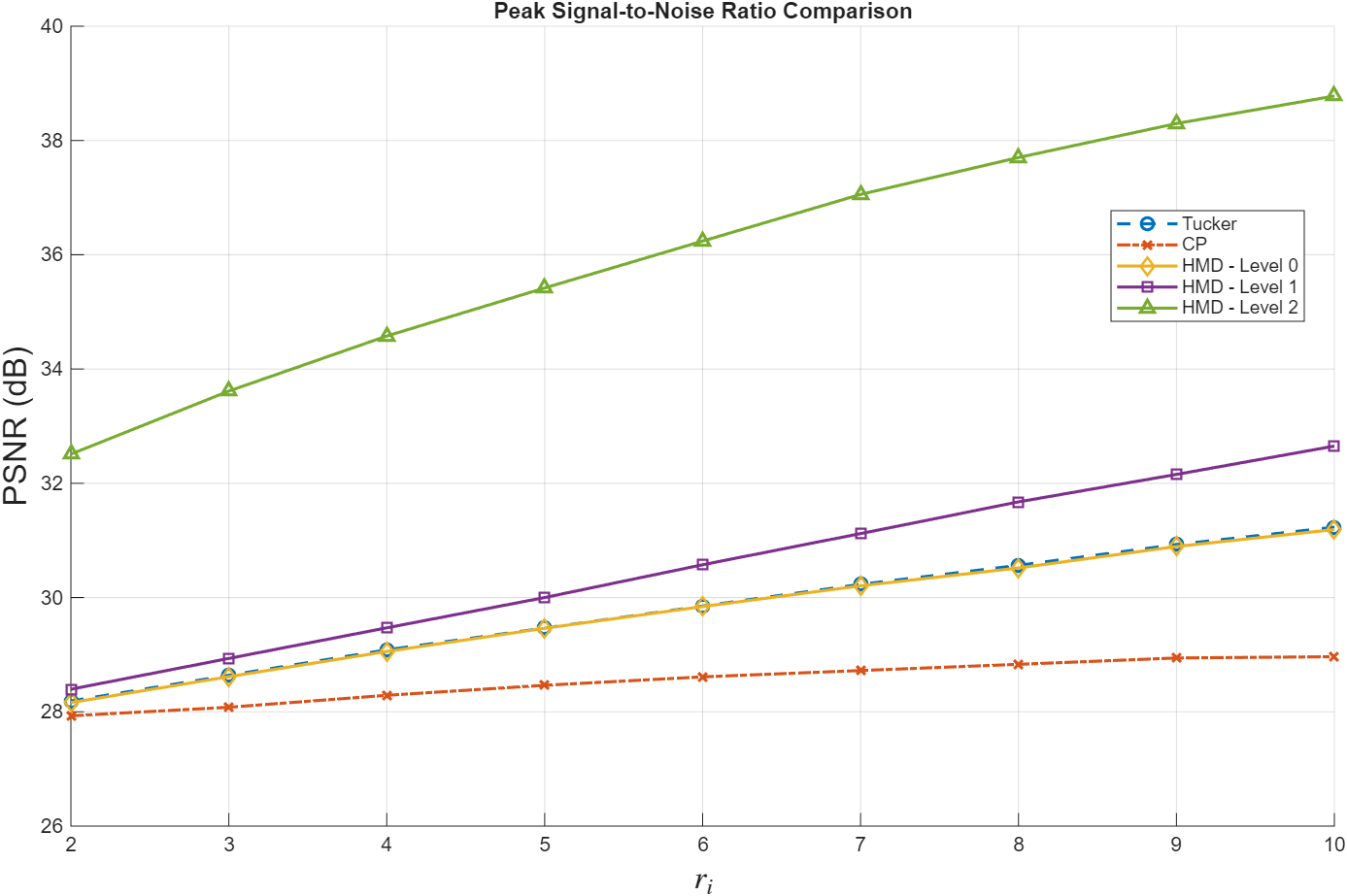}
    \caption{PSNR (dB) performance across varying ranks $r_i$ for the KTH video dataset}
    \label{fig:Video_PSNR}
\end{figure}
%------------------Fig8------------------------

Analyzing the PSNR in Fig. \ref{fig:Video_PSNR} further emphasizes the robust signal fidelity of the proposed framework. In video processing, maintaining high signal quality requires effectively isolating active foreground movement from static background noise under tight compression constraints. While CP, Tucker, and the lower level HMDs cluster at the lower performance boundaries, the second level HMD completely redefines the experiment's limits. Because its mathematical structure isolates coupled spat-temporal interactions while deflating independent background variance, it inherently suppresses noise. This algebraic advantage translates to an exceptional initial fidelity of roughly $32.5$ dB at a minimal rank of $r_i = 2$, soaring to nearly $39$ dB at $r_i = 10$.

%------------------Fig9------------------------
\begin{figure}[h!]
  \centering
    \includegraphics[scale=.5]{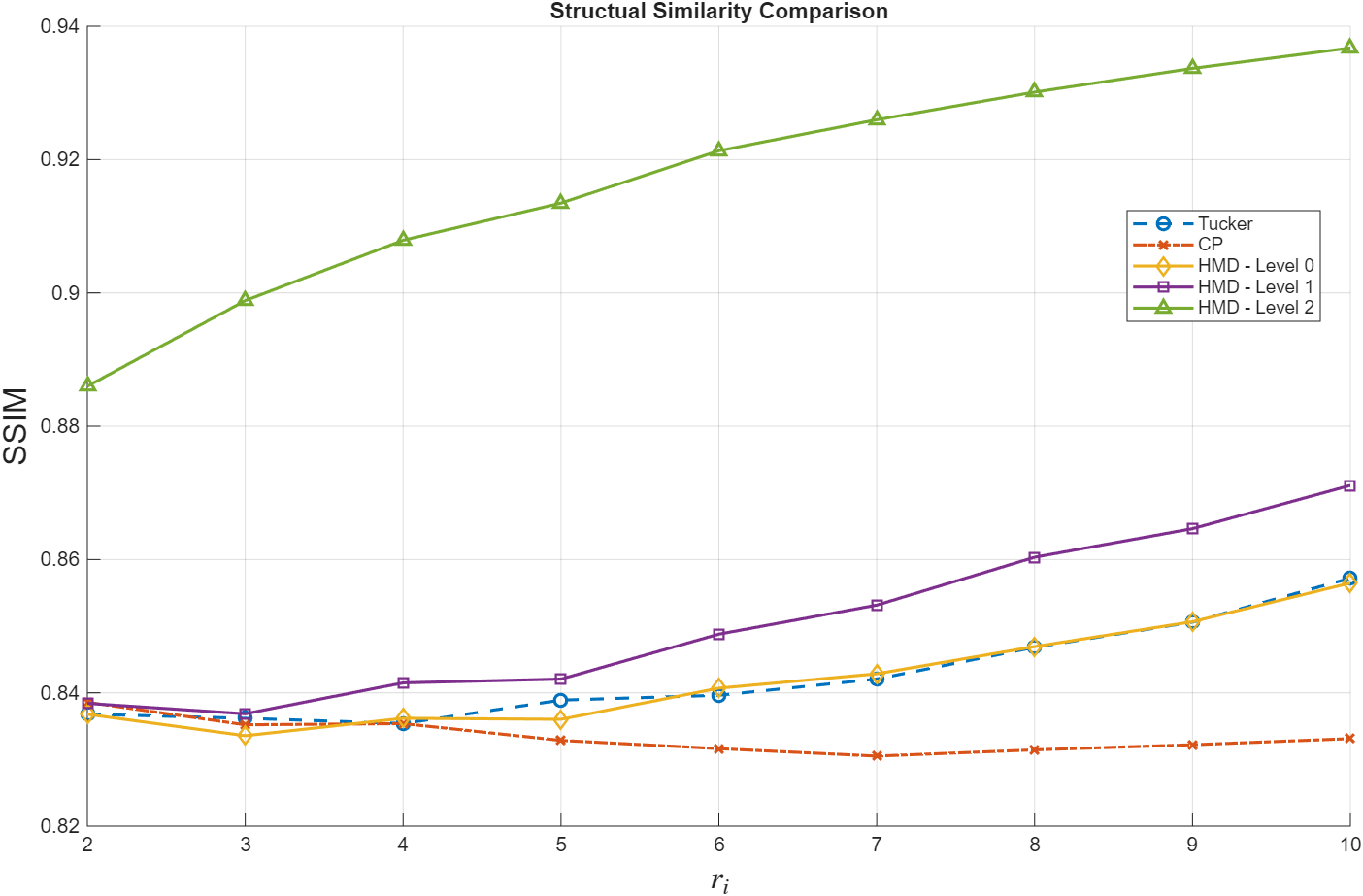}
    \caption{SSIM structural evaluation under different rank configurations $r_i$ for the KTH video dataset}
    \label{fig:Video_SSIM}
\end{figure}
%------------------Fig9------------------------

SSIM metric in Fig. \ref{fig:Video_SSIM} provides critical insight into the preservation of the video's spatio-temporal geometry. Retaining structural boundaries across consecutive frames is paramount to preventing the blurring artifacts typical of traditional low-rank methods. While the classical models struggle or even experience slight structural degradation at lower ranks, the second level HMD immediately secures a commanding advantage. Starting with a highly intact SSIM of approximately $0.89$ at $r_i = 2$, it climbs smoothly to nearly $0.94$ at $r_i = 10$. Geometrically, this confirms that mapping the two-dimensional relational manifolds between spatial structures and temporal shifts ensures the reconstructed action remains strictly faithful to the original recording.

%------------------Fig10-----------------------
\begin{figure}[h!]
  \centering
    \includegraphics[scale=.5]{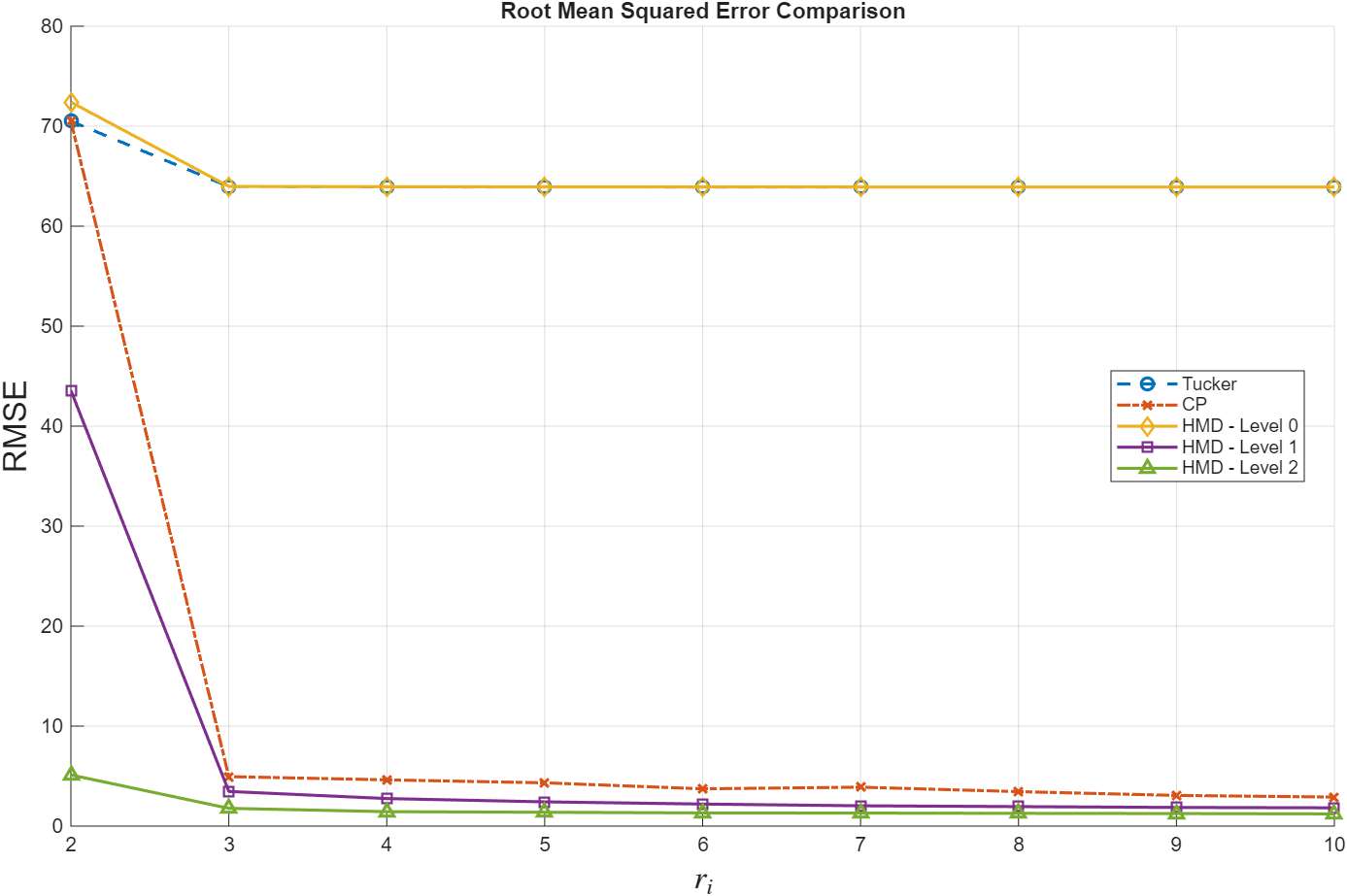}
    \caption{RMSE comparison as a function of rank $r_i$ for the amino acid dataset}
    \label{fig:Amino_RMSE}
\end{figure}
%------------------Fig10-----------------------
In the final experiment, the RMSE evaluation for the amino acid dataset \citep{aminoData} presents a structural challenge. To this end, we deliberately constrained the first mode's rank ($r_1$) to $2$, while varying the remaining ranks ($r_i$) from $2$ to $10$. Because the dataset is composed of exactly three chemical compounds, this rank-$2$ bottleneck prevents classical CP and Tucker methods from ever resolving the independent signatures, trapping their errors at high plateaus. This is where the second-level HMD truly shines. By explicitly capturing the coupled excitation-emission interactions between the remaining dimensions, it seamlessly bypasses this mathematical limitation. Rather than being restricted by the deficient concentration mode, it successfully maps the underlying spectroscopic fingerprints to minimize residual error. Ultimately, the second-level HMD achieves an accurate, physical reconstruction that traditional models simply cannot match under the same tight constraints.

\section{Concluding Remarks}
\label{sec:conc}

In this paper, we introduced a novel tensor decomposition method named Holistic Multivariance Decomposition (HMD). The proposed framework operates as an adjustable low-rank approximation technique similar to the Tucker decomposition, while simultaneously capturing complex cross-mode interrelations in a manner inspired by EMPR. By uniquely addressing both challenges within a unified algebraic structure, HMD achieves exceptional success in reconstructing multivariate datasets across distinct domains.

Architecturally, HMD adopts multiple core tensors whose individual parameter counts remain significantly smaller than that of the original tensor, and it utilizes support matrices with an adjustable number of columns. This inherent structural flexibility allows researchers to  tune the number of parameters included in the representation, enabling them to achieve the desired balance between computational complexity and high reconstruction quality.

Our numerical experiments demonstrate that HMD provides a robust and dependable tensor decomposition framework. When evaluated on benchmark applications spanning hyperspectral imagery, dynamic video analysis, and chemometrics, HMD consistently yields superior reconstruction accuracy and signal fidelity, even when the factor (support) matrices are constrained to fixed low ranks.

The selection of these support matrices remains a crucial step for tailoring HMD to specific data characteristics and achieving exclusive modeling success. Consequently, any contextually relevant semi-orthogonal matrix set, such as incomplete Fourier or wavelet bases, can be seamlessly integrated into the framework to further enhance reconstruction quality. 

Given its versatile attributes, structural flexibility, and strong numerical performance, the proposed HMD framework emerges as a highly competitive and efficient tensor decomposition alternative for resolving complex multidimensional data structures across science and engineering.

% --- References ---
% Loading bibliography database
\bibliographystyle{plainnat}
\bibliography{references}

@article{tunaTGRS,
  title={Iterative enhanced multivariance products representation for effective compression of hyperspectral images},
  author={Tuna, S{\"u}ha and T{\"o}reyin, Beh{\c{c}}et U{\u{g}}ur and Demiralp, Metin and Ren, Jinchang and Zhao, Huimin and Marshall, Stephen},
  journal={IEEE Transactions on Geoscience and Remote Sensing},
  volume={59},
  number={11},
  pages={9569--9584},
  year={2020},
  publisher={IEEE}
}

@article{tunaTCBB,
  title={Gene Teams are on the Field: Evaluation of Variants in Gene-Networks Using High Dimensional Modelling},
  author={Tuna, Suha and Gulec, Cagri and Yucesan, Emrah and Cirakoglu, Ayse and Arguden, Yelda Tarkan},
  journal={IEEE/ACM Transactions on Computational Biology and Bioinformatics},
  volume={20},
  number={5},
  pages={2959--2969},
  year={2023},
  publisher={IEEE}
}

@article{tunaFranklin,
  title={A new feature extraction scheme based on support optimization in Enhanced Multivariance Products Representation for Hyperspectral Imagery},
  author={{\c{S}}en, Muhammed Enis and Tuna, S{\"u}ha},
  journal={Journal of the Franklin Institute},
  volume={362},
  number={2},
  pages={107464},
  year={2025},
  publisher={Elsevier}
}

@inproceedings{tunaCNN,
  title={Compression of Convolutional Neural Networks Employing Tensor Train and High Dimensional Model Representation},
  author={Y{\i}lmaz, Berna and Tuna, S{\"u}ha},
  booktitle={2025 9th International Symposium on Innovative Approaches in Smart Technologies (ISAS)},
  pages={1--6},
  year={2025},
  organization={IEEE}
}

@article{koldaTensor,
  title={Tensor decompositions and applications},
  author={Kolda, Tamara G and Bader, Brett W},
  journal={SIAM review},
  volume={51},
  number={3},
  pages={455--500},
  year={2009},
  publisher={SIAM}
}

@techreport{koldaMultilinear,
  title={Multilinear operators for higher-order decompositions.},
  author={Kolda, Tamara Gibson},
  year={2006},
  institution={Sandia National Laboratories}
}

@article{hsiTensor,
  title={Tensor decompositions for hyperspectral data processing in remote sensing: A comprehensive review},
  author={Wang, Minghua and Hong, Danfeng and Han, Zhu and Li, Jiaxin and Yao, Jing and Gao, Lianru and Zhang, Bing and Chanussot, Jocelyn},
  journal={IEEE Geoscience and Remote Sensing Magazine},
  volume={11},
  number={1},
  pages={26--72},
  year={2023},
  publisher={IEEE}
}

@article{chemoTensor,
  title={A new penalized nonnegative third-order tensor decomposition using a block coordinate proximal gradient approach: Application to 3D fluorescence spectroscopy},
  author={Vu, Xuan and Chaux, Caroline and Thirion-Moreau, Nad{\`e}ge and Maire, Sylvain and Carstea, Elfrida Mihaela},
  journal={Journal of Chemometrics},
  volume={31},
  number={4},
  pages={e2859},
  year={2017},
  publisher={Wiley Online Library}
}

@article{low_rankreview,
  title={A review on low-rank models in data analysis},
  author={Lin, Zhouchen},
  journal={Big Data \& Information Analytics},
  volume={1},
  number={2/3},
  pages={139--161},
  year={2016},
  publisher={Big Data \& Information Analytics}
}

@article{tunaHDMR,
  title={An efficient feature extraction approach for hyperspectral images using Wavelet High Dimensional Model Representation},
  author={Tuna, S{\"u}ha and Korkmaz {\"O}zay, Evrim and Tunga, Burcu and G{\"u}rvit, Ercan and Tunga, M Alper},
  journal={International Journal of Remote Sensing},
  volume={43},
  number={19-24},
  pages={6899--6920},
  year={2022},
  publisher={Taylor \& Francis}
}

@inproceedings{sota_3dspeck,
  title={Hyperspectral image compression with modified 3D SPECK},
  author={Ngadiran, Ruzelita and Boussakta, Said and Bouridane, Ahmed and Syarif, Bayan},
  booktitle={2010 7th International Symposium on Communication Systems, Networks \& Digital Signal Processing (CSNDSP 2010)},
  pages={806--810},
  year={2010},
  organization={IEEE}
}

@article{signalTensor,
  title={Tensor decomposition for signal processing and machine learning},
  author={Sidiropoulos, Nicholas D and De Lathauwer, Lieven and Fu, Xiao and Huang, Kejun and Papalexakis, Evangelos E and Faloutsos, Christos},
  journal={IEEE Transactions on signal processing},
  volume={65},
  number={13},
  pages={3551--3582},
  year={2017},
  publisher={IEEE}
}

@article{biomedicalTensor,
  title={Tensor methods in biomedical image analysis},
  author={Sedighin, Farnaz},
  journal={Journal of Medical Signals \& Sensors},
  volume={14},
  number={6},
  pages={16},
  year={2024},
  publisher={Medknow}
}

@inproceedings{tensor2d,
  title={A survey of tensor methods},
  author={De Lathauwer, Lieven},
  booktitle={2009 IEEE international symposium on circuits and systems},
  pages={2773--2776},
  year={2009},
  organization={IEEE}
}

@INPROCEEDINGS{videoData,
  author={Schuldt, C. and Laptev, I. and Caputo, B.},
  booktitle={Proceedings of the 17th International Conference on Pattern Recognition, 2004. ICPR 2004.},
  title={Recognizing human actions: a local SVM approach},
  year={2004},
  volume={3},
  pages={32-36},
  doi={10.1109/ICPR.2004.1334462}
}

@article{aminoData,
  author    = {Bro, Rasmus},
  title     = {{PARAFAC}: Tutorial and applications},
  journal   = {Chemometrics and Intelligent Laboratory Systems},
  volume    = {38},
  number    = {2},
  pages     = {149--171},
  year      = {1997},
  publisher = {Elsevier},
  doi       = {10.1016/S0169-7439(97)00032-4}
}

@misc{indianData,
  title = {220 Band {AVIRIS} Hyperspectral Image Data Set: June 12, 1992 Indian Pine Test Site 3},
  author = {Baumgardner, Marion F. and Biehl, Larry L. and Landgrebe, David A.},
  year = {1995},
  publisher = {Purdue University},
  doi = {10.4231/R7RX991C},
  url = {https://purr.purdue.edu/publications/1947/1}
}

@inproceedings{burcuEMPR,
  title={An Iterative Enhanced Multivariance Product Representation Scheme for Multivariate Function Decomposition},
  author={Tunga, B and Demiralp, M},
  booktitle={Proceedings of the International Conference on Applied Computer Science},
  volume={1},
  pages={247--253},
  year={2010}
}

@article{burcuHDMR,
  title={Constancy maximization based weight optimization in high dimensional model representation for multivariate functions},
  author={Tunga, Burcu and Demiralp, Metin},
  journal={Journal of mathematical chemistry},
  volume={49},
  number={9},
  pages={1996--2012},
  year={2011},
  publisher={Springer}
}

\end{document}